\documentclass{emulateapj}
\usepackage{amsmath}

\shorttitle{Photoevaporation of Protoplanetary Disks}
\shortauthors{Balog et al.}

\begin{document}

\title{Photoevaporation of Protoplanetary Disks}

\author{Zoltan Balog\altaffilmark{1}, George H. Rieke, James Muzerolle}
\affil{Steward Observatory, University of Arizona, 933 N. Cherry Ave. Tucson, AZ, 85721}
\email{zbalog@as.arizona.edu}

\and

\author{John Bally}
\affil{Center for Astrophysics and Space Astronomy, University of Colorado, 389 UCB, Boulder, CO 80309}

\and 

\author{Kate Y. L. Su, Karl Misselt, Andr\'as G\'asp\'ar}
\affil{Steward Observatory, University of Arizona, 933 N. Cherry Ave. Tucson, AZ, 85721}

\altaffiltext{1}{on leave from Dept of Optics and Quantum Electronics, University of Szeged, H-6720, Szeged, Hungary}

\begin{abstract}
We present HST/NICMOS Paschen $\alpha$ (Pa$\alpha$) images and low and high resolution IRS spectra of photoevaporating disk-tail systems originally detected at 24$\mu$m near O stars. We find no Pa$\alpha$ emission in any of the systems. The resulting upper limits correspond to about 2$-$3 $\times$ 10$^{-6}$ M$_{\odot}$ of mass in hydrogen in the tails suggesting that the gas is severely depleted. The IRAC data and the low resolution 5$-$12 $\mu$m IRS spectra provide evidence for an inner disk while high resolution long wavelength (14$-$30 $\mu$m) IRS spectra confirm the presence of a gas free ``tail'' that consists  of $\sim$ 0.01 to $\sim$ 1 $\mu$m dust grains originating in the outer parts of the circumstellar disks. Overall our observations support theoretical predictions in which photoevaporation removes the gas relatively quickly ($\leq 10^5$ yrs) from the outer region of a protoplanetary disk but leaves an inner more robust and possibly gas-rich disk component of radius 5-10 AU. With the gas gone, larger solid bodies in the outer disk can experience a high rate of collisions and produce elevated amounts of dust. This dust is being stripped from the system by the photon pressure of the O star to form a gas-free dusty tail.

\end{abstract}

 
\section{Introduction}

 The timescale at which a protoplanetary disk loses its primordial gas, which dominates the total disk mass, is a crucial factor regarding the outcome of the planet forming process in the vicinity of high mass stars. The photoevaporation of protoplanetary disks by external UV radiation in an environment dominated by high mass O-stars is a particularly potent mechanism for removing disk material. The timescale of photoevaporation is governed by the amount of UV radiation from the nearby O-stars, which significantly erodes the protoplanetary disk in less than a million years. Studies of disk erosion agree that the outer part of the disk evaporates quickly; however according to, e.g., \citet{Holl04} and \citet{Adam04} the erosion of the disk slows down significantly when the size reaches 10 AU. Other authors \citep[e.g.][]{John98,Mats03b} have found that the photoevaporation process can shrink the radius of the disk below 1 AU within $10^6$ years and it can even destroy the whole disk when it is coupled with internal viscous evolution. Therefore, the presence of high mass stars might be lethal for forming planets since the EUV and FUV radiation of a nearby O-star can disrupt the protoplanetary disk on a short time scale (a couple of times $10^5$ yr) and stop the planet formation process. On the other hand the radiation might trigger the formation of rocky planets by accelerating the dust settling and grain growth in the midplane of the disk \citep{Thro05}. 


\citet{Balo06} reported the discovery of three comet like structures near different O-type stars (with the comet-like tails pointing away from the central high mass star) in 24 $\mu$m {\it Spitzer}/MIPS images. Adopting a radiation-pressure driven outflow disk model \citep{Su05} they concluded that these objects are photoevaporating protoplanetary disks where dust entrained in the photoevaporative flow is responsible for the mid-infrared tails detected at 24 $\mu$m. The tails extend up to 0.2 pc from the evaporation working surface, so they are convenient for detailed study.  To investigate this phenomenon further and to get a better understanding of the nature of these objects, we have searched these three systems for the gas expected to be a significant component of the flow. We used Pa$\alpha$ as an ideal tracer of the photoevaporation process, since it is the strongest emission line in the near infrared and is not strongly affected by extinction. We also report mid-IR spectra to provide a better understanding of the physical processes in the disk and to help determine the properties of the dust.

\section{Observations}

\subsection{HST/NICMOS}
We used HST/NICMOS to search for ionized hydrogen, predicted to be an abundant product of photoevaporation in a protoplanetary disk \citep{Rich00}. We chose the filters F187N and F190N (Pa$\alpha$ line and a nearby lineless continuum) because we expected to detect the strongest emission in these filters. Our targets were the three objects detected at 24 $\mu$m by \citet{Balo06}. All these objects have a cometary structure in the 24 $\mu$m {\it Spitzer}/MIPS images with the ``comets' tail'' pointing way from a nearby O star (see Fig. 1 of \citet{Balo06}).

The exposure time for both the line and the continuum filters was 272 s, 352 s and 671 s in the case of objects in NGC~2244, NGC~2264 and IC~1396 respectively. We employed scale factors of 3.2641699$\times 10^{-5}~{\rm Jy~s~DN^{-1}}$ for F190N and 3.2329899$\times 10^{-5}~{\rm Jy~s~DN^{-1}}$ for F187N in the case of NGC~2264, 3.2536798$\times 10^{-5}~{\rm Jy~s~DN^{-1}}$ for F190N and 3.1735399$\times 10^{-5}~{\rm Jy~s~DN^{-1}}$ for F187N in the case of NGC~2244 and 3.2641699$\times 10^{-5}~{\rm Jy~s~DN^{-1}}$ for F190N and  \\3.2329899$\times 10^{-5}~ {\rm Jy~s~DN^{-1}}$ for F187N in the case of IC~1396 to convert the original frames to calibrated flux images. 

\begin{figure}
\plottwo{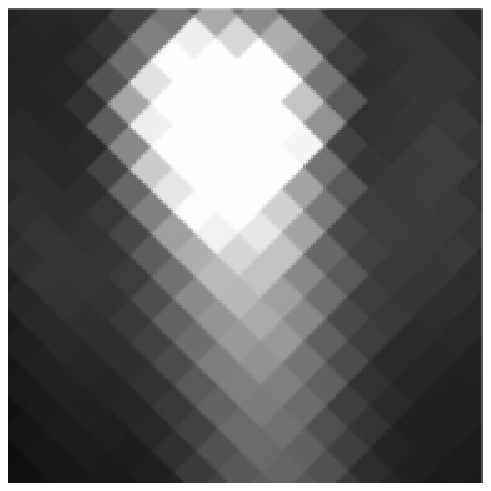}{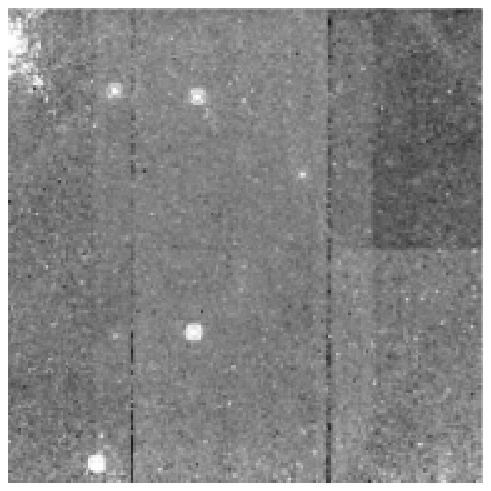}
\plottwo{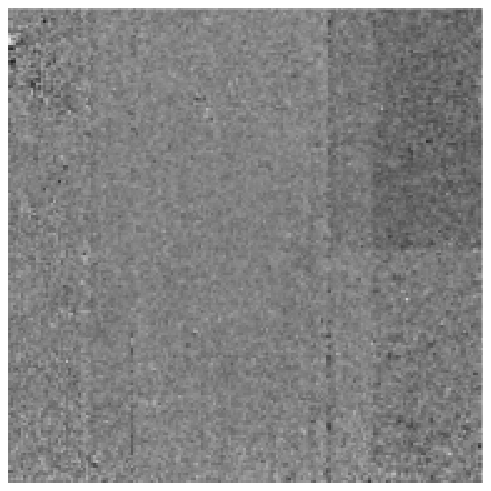}{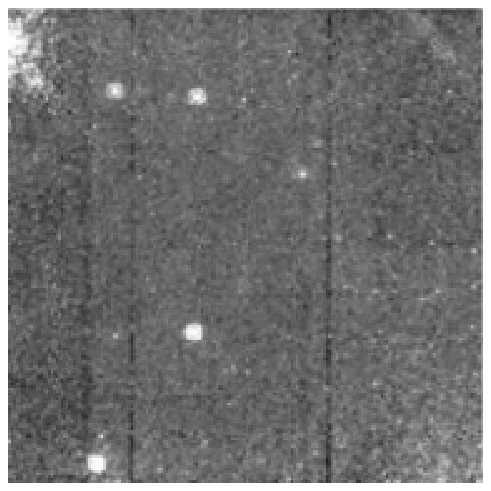}
\caption{The images of the photoevaporating source in NGC~2244. Upper left: 24 $\mu$m; upper right Pa$\alpha$; lower right: Pa$\alpha$ continuum; lower left continuum subtracted Pa$\alpha$.  The covered area is $19\arcsec \times 19\arcsec$ (0.138 $\times$ 0.138 pc in the distance of NGC2244) in all four panels.}
\label{fig:2244}
\end{figure}

As an example Fig. \ref{fig:2244} shows the source in NGC~2244 at 24 $\mu$m, at 1.87 $\mu$m ( Pa$\alpha$ line) and in the continuum subtracted Pa$\alpha$ images. The covered area is $19\arcsec \times 19\arcsec$ (0.138 $\times$ 0.138 pc in the distance of NGC2244). The other two sources behave similarly. It is clear that the tail structure is not visible in the Pa$\alpha$ images even after continuum subtraction. However in all cases there is an unresolved point source in the comets' head, which confirms that the photoevaporating source is indeed a star and not an extended globule.

We performed aperture photometry on the continuum images following the prescription in the NICMOS data handbook to estimate the continuum flux of the central object. In two cases (NGC~2264 and IC~1396) the NICMOS fluxes are in good agreement with fluxes calculated from the 2MASS data (See Figs. \ref{fig:1396IRSSL} and \ref{fig:2264IRSSL}). For NGC~2244 the nearby O star strongly affects the 2MASS magnitudes so the fluxes calculated from the 2MASS data are much higher than the flux we obtained using NICMOS. However VLT/VIMOS R and I fluxes (Robert King private communication) confirm that our NICMOS measurements are accurate and the 2MASS data are about 1-2 magnitudes higher than the real values. Therefore we ignore the 2MASS data for this source and use only the R, I and NICMOS 1.9 $\mu$m data to estimate the SED of the central object (Fig. \ref{fig:2244IRSSL}).  The reason why this problem does not affect the other two objects is that they are about five times farther (in angular distance) from the O star than the source in NGC2244.

\subsection{{\it Spitzer}/IRS}

\begin{deluxetable*}{lccccc}
\tablecolumns{6}
\tablewidth{0pc}
\tablecaption{The {\it Spitzer}/IRS observations of the photoevaporating tails.}
\tablehead{
\colhead{Cluster}&\colhead{J2000} & &\colhead{Integration} \\
 & &\colhead{SL1} & \colhead{SL2} &\colhead{SH} & \colhead{LH} \\
&\colhead{[h:m:s] [d:m:s]} & \colhead{[s]} & \colhead{[s]}  & \colhead{[s]}  & \colhead{[s]}
}
\startdata
NGC~2244 & 06:31:54.68 04:56:25.0& 18.1 & 73.4& 25.1& 31.5 \\
IC~1396  & 21:38:57.09 57:30.46.5& 146.8& 146.8& 487.5& 2176.4 \\
NGC~2264 & 06:41:01.92 09:52:39.0& 146.8 & 146.8& 125.8& 457.6 \\
\enddata
\end{deluxetable*}

We used Spitzer/IRS to take low and high resolution mid-IR spectra of the photoevaporating sources to explore the physical processes in the evaporating gas. All spectra were taken centered at the ``head'' positions. The exposure times for the four IRS modules (short-low 1st order (SL1), short-low 2nd order (SL2), short-high (SH), and long-high (LH)) are given in Table 1. We extracted the low resolution spectra from the Spitzer Science Center (SSC) calibrated post-bcd background subtracted frames using SPICE\footnote{Spitzer IRS Custom Extraction (SPICE) is a JAVA-based tool for interactively extracting {\it Spitzer}/IRS spectra.}. We averaged the two nod positions to increase the S/N. For the high resolution modes we subtracted a spectrum obtained at a nearby location before we extracted the science spectra, to eliminate contamination from the surrounding HII region. We also discarded the beginning and the end of each order. We calibrated the extracted spectra using the 5.8$-$24 $\mu$m photometry. No emission lines are apparent in the high resolution spectra, indicating the absence of gas. However, the Pa$\alpha$ observations place more stringent limits on the amount of gas. We smoothed the IRS high resolution spectra to reduce the noise, and then merged them to get full coverage between 10 and 30 $\mu$m.

\begin{figure}
\plotone{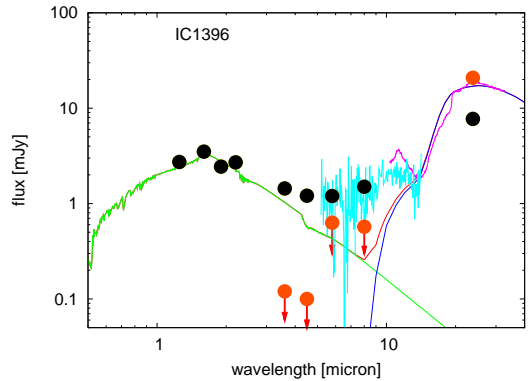}
\caption{Spectrum of the head of the photoevaporating source in IC~1396 (light blue line: low resolution; magenta line: smoothed high resolution). The green line represents the continuum of a star with stellar temperature of 4000 K and blue line represents the SED of our dust tail model while the red line is the sum of the star and tail models. Black dots show the fluxes of the central point source in different photometric bands, orange dots show the fluxes of the tail (we give 3$\sigma$ upper limits where the tail is not detected)}
\label{fig:1396IRSSL}
\end{figure}

\begin{figure}
\plotone{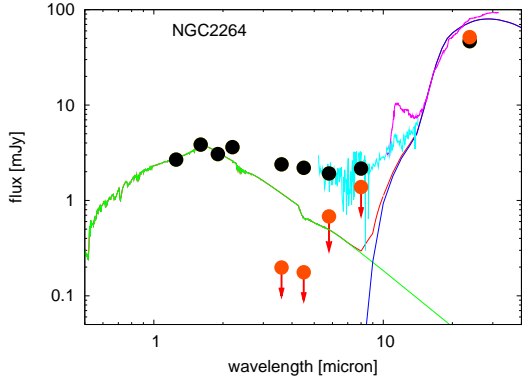}
\caption{Same as in Fig \ref{fig:1396IRSSL}, for the source in NGC~2264}
\label{fig:2264IRSSL}
\end{figure}

\begin{figure}
\plotone{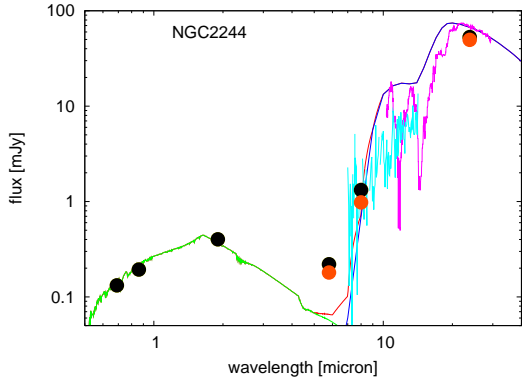}
\caption{Same as in Fig \ref{fig:1396IRSSL}, for the source in NGC~2244. We note that the low S/N in the short wavelength portion of the high resolution spectrum resulted in two spurious absorption-like features in the final smoothed spectrum. These features are not real so we do not discuss them any further.}
\label{fig:2244IRSSL}
\end{figure}

\begin{figure}
\plottwo{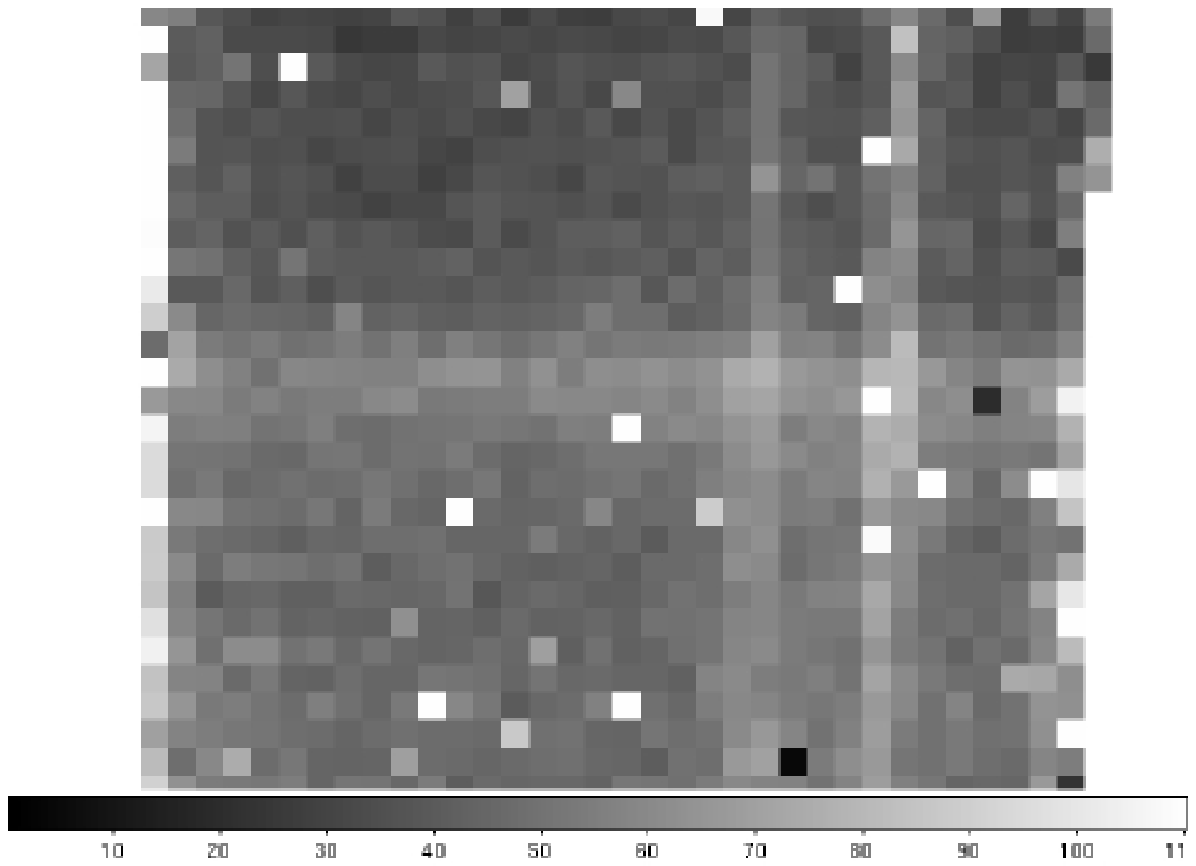}{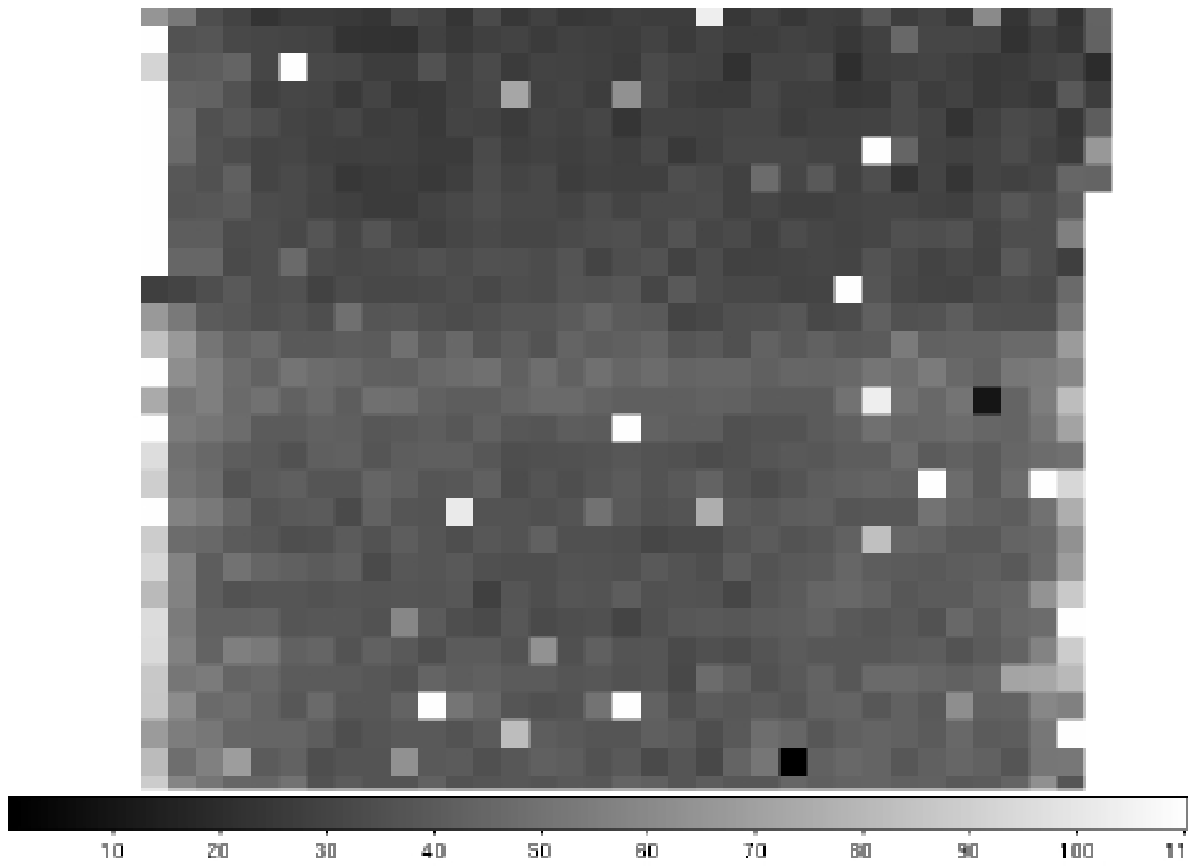}
\caption{Part of the low resolution spectrum of the source in NGC~2264 (left panel) and the same part of a low resolution spectrum taken at the nearby background location where the high resolution background spectra were taken (right panel). The exposure times for the two spectra are exactly the same as well as the stretch of the images as indicated by the greyscale code at the bottom of each image. The direction of the dispersion is along the vertical axis while the horizontal axis represents the spatial direction along the slit.}
\label{fig:compPAH}
\end{figure}

We show our spectra in Figs. \ref{fig:1396IRSSL}, \ref{fig:2264IRSSL}, and \ref{fig:2244IRSSL} along with the Kurucz model spectra for the stellar photospheres and our model SEDs of unbound small particles. The IR excess is clearly visible even in the unsmoothed low S/N low resolution spectra. The high resolution spectra show a bump around 11.3 $\mu$m that is not detected in the low resolution spectra. This feature originates from aromatic emission and is very common in HII regions. The reason we detect it at high resolution and do not detect it at low resolution can be explained by the difference in the background subtraction in the two observing modes. In the short-low (SL) mode we subtract the two nod positions from each other to eliminate the background while in high resolution mode a different spectrum taken at a nearby location was used for the same purpose. We see the 11.3 $\mu$m line in the latter case as a result of imperfect background subtraction due to spatial variation of the line flux. To support this explanation we show the relevant part of the low resolution spectrum of the source in NGC~2264 (Fig. \ref{fig:compPAH} left panel) and the same part of a low resolution spectrum taken at the nearby background location where the high resolution background spectra were taken (Fig. \ref{fig:compPAH} right panel). The two sharp vertical lines are the spectra of our target and a nearby star that happened to fall on the slit while the smeared out horizontal line in the middle of the image is the 11.3 $\mu$m aromatic feature. The facts that the aromatic feature is visible through the entire slit width and that it is in both spectra suggest that it comes from the background rather than from the source itself. It is also clear that the feature is much weaker at the background location (right panel) than at the source position (left panel).   

\subsection{Archival {\it Spitzer}/IRAC-MIPS data}
We downloaded all the available 3.6-8 $\mu$m data from the {\it Spitzer} archive to see if new observations have become available since the appearance of the paper reporting the discovery of these objects \citep{Balo06}. Only in the case of NGC~2244 did we find new deeper IRAC observations (Bouwman et al., Prog ID: 30726). For the rest of the sources we used the data from the Prog ID: 37 (PI: Fazio) and Prog ID: 58 (PI: Rieke) GTO programs.

We re-mosaiced all the IRAC observations using custom IDL scripts and extracted the flux at the position of the ``head'' of each source using 2 pixel apertures. We estimated 3$\sigma$ upper limits for the flux of the ``tail'' in the case of NGC~2264 and IC~1396. The situation for NGC~2244 was more complicated since the nearby O star affects the photometry at shorter wavelengths. As a result, although we detect the source at 3.6 and 4.5 $\mu$m we cannot extract its flux. At 5.8 and 8.0 $\mu$m both the ``head'' and the ``tail'' are measurable in the new mosaics.

For the 24 $\mu$m  MIPS photometry we used the mosaics originally generated with the DAT \citep{Gordon05} and published by \citet{Sici06}, \citet{Teix06}, and \citet{Balo07}. First we measured the flux of the ``head'' via PSF fitting using IDP3\footnote{Image Display Paradigm 3 (IDP3) is an IDL (Interactive Data Language) package for analyzing astronomical images developed by the NICMOS Software Group at the University of Arizona.} then we estimated the flux in the tail. To do so, we selected a rectangular region along the tail in the PSF subtracted image and measured the signal in this region. Next we moved the selected region to nearby background positions on both sides of the tail and averaged the signal measured at these positions. We then subtracted the averaged background from the signal at the position of the tail to get the total flux in the tail. The results of our photometry are plotted in Figs.  \ref{fig:1396IRSSL}, \ref{fig:2264IRSSL}, and \ref{fig:2244IRSSL} (black dots: head, orange dots: tail).

\section{Results}
\subsection{Calculation of Hydrogen Mass Upper Limit}

We detected no Pa$\alpha$ emission in any of the objects; we will now derive upper limits to the amount of gas compatible with this result.

\subsubsection{Photoevaporation from an Optically Thick (to Lyman continuum) Disk}

Consider a spherical (or hemispherical) cloud of hydrogen with a radius $r$, located at a distance $d$ from an O star that has a Lyman continuum luminosity $Q$ (in ${\rm \gamma s^{-1}}$). If the cloud is sufficiently dense, it will be optically thick to the Lyman continuum radiation. In the model developed for the Orion Nebula proplyds \citep{John98}, soft, non-ionizing UV heats the outer layers of the cloud resulting in a quasi-steady-state, soft-UV driven wind expanding with a velocity of a few kilometers per second. Ionizing Lyman continuum radiation forms a quasi-stationary ionization front (I-front) in this wind where the gas becomes ionized and heated to about $10^4$ K. The resulting pressure jump drives a D-type shock into the expanding low-velocity neutral wind that compresses and decelerates the flow before it passes through the I-front. At the I-front, the plasma accelerates to about the sound speed in the ionized gas, c$_{II}$, and the pressure gradient results in a spherically divergent flow. Once ionization equilibrium has been established in the ionized flow, and the system is in a steady state of photo-ablation driven mass loss, the flux of Lyman continuum incident on the cloud, $F = Q/4 \pi d^2$ is balanced by the recombination rate between the I-front and the illuminating star. For a constant velocity, spherically divergent flow, $n_e(r) = n_I(r_I/r)^2$ where $n_I$ is the plasma density just outside (on the irradiated side) of the ionization front, and $r_I$ is the ionization front radius. The total recombination rate in the column between the O star and the I-front is then given by

\begin{equation}
F = {{Q} \over{4 \pi d^2}} = \int {n^2_e (r)\alpha_{B}dr} = {{n^2_I \alpha_B r_I} \over 3}
\end{equation}

\noindent where the integration limits run from $r_I$ to infinity and $\alpha_B \approx 2.6 \times 10^{-13} (cm^3 s^{-1})$ is the case-B recombination coefficient for hydrogen at $10^4$ K. For a disk or circumstellar environment with a mean neutral gas density $n(H) >> (3F/\alpha_B r_I)^{1/2}$ inside radius $r_I$, the electron density at the I-front is self-regulated by the Lyman continuum flux to be $n_I$. A higher electron density would shut off the UV, decrease the mass loss, and therefore lower the density. A lower electron density would result in a higher flux at the I-front and a greater mass-loss rate until the density at the I-front reached the value $n_I$ . If the mean neutral gas density is lower than $n_I$, the I-front would fully ionize the circumstellar environment which would then be optically thin to Lyman continuum.

This simple model ignores the expected increase in velocity of the photo-ablating plasma due to the density and pressure gradient. In a full treatment of the hydrodynamics, the flow velocity will increase to about 2 to 3 $\times c_{II}$ as the density decreases by one to two orders of magnitude. However this acceleration will have little effect on the observed surface brightness of species such as Pa$\alpha$ because most of the line emission is produced by the dense plasma at the base of the ionized flow, since line intensities are proportional to the emission measure.

The emission measure is defined in the usual manner to be $EM = \int n^2_e(l)dl$ where $n_e$ is the electron density and $dl$ is an infinitesimal length element along the line-of-sight. For a hemispherical or spherical surface giving rise to a constant velocity wind with an $r^{-2}$ density profile, the highest EM occurs along a line-of-sight that just grazes the ionization front. By integrating the line-of-sight along a tangent to the surface, it can be shown that
$EM = (\pi/2[pc])n^2_I r_I$ where $pc = 3.086 \times 10^{18} {\rm cm}$ converts the units of EM from c.g.s. to $\rm {cm^{-6} pc}$. From ionization equilibrium, $n^2_I r_I \approx 3Q/4\pi \alpha_B d^2$, independent of the ionization front radius, so that $EM \approx 3Q/8[pc]\alpha_B d^2$. The approximate equality is used here because we ignore the slight acceleration of the ionized front due to the pressure gradient. Thus, the expected surface brightness of any emission line is, to first order, independent of the cloud radius and inversely dependent on the square of the distance from the source of ionization.

The flux of any recombination line of hydrogen can be converted into an estimate of the emission measure. Following \citet{Spit78}, the intensity of a transition from level m to level n is given by

\begin{equation}
\begin{split}
I_{mn}= \int j_{\nu} d\nu = {{h\nu_{mn}\alpha_{mn}} \over {4\pi}} \int n^2_e(l)dl = \\
2.46 \times 10^{17} h\nu_{mn}\alpha_{mn} EM
\end{split}
\end{equation}

\noindent where $\alpha_{mn}$ is the total recombination into the upper level of the transition taking account of all recombinations into higher levels that cascade into the level of interest and of the branching ratio for transitions to lower levels. EM is the emission measure in units of ${\rm cm^{-6} pc}$. For full ionization, $n_e = n_p = n(H)$ is assumed. The intensities of various hydrogen recombination lines relative to H$\beta$ were tabulated by \citet{Peng64}. From his tables, the Case-B recombination coefficient for Pa$\alpha$ is $\alpha_{43} = 4.10 \times 10^{-14} ({\rm cm^3 s^{-1}})$. Thus, the relationship between Pa$\alpha$ flux  in units of $\rm {erg s^{-1} cm^{-2} arcsec^{-2}}$ and EM is $F_{P\alpha} = 2.51 \times 10^{-19} EM$ where EM is in units of ${\rm cm^{-6}pc}$. 

\begin{deluxetable*}{lccc}
\tablecolumns{4}
\tablewidth{0pc}
\tablecaption{The HST/NICMOS Pa$\alpha$ observations of the photoevaporating tails.}
\tablehead{
\colhead{Cluster}&\colhead{J2000} & \colhead{Integration} & \colhead{$F_{P\alpha}$ limit}\\
&\colhead{[h:m:s] [d:m:s]} & \colhead{[s]} & \colhead{[$\rm erg~cm^{-2}s^{-1}arcsec^{-2}$] }
}
\startdata
NGC~2244 & 06:31:54.68 04:56:25.0& 272 & $4.2\times10^{-16}$ \\
IC~1396  & 21:38:57.09 57:30.46.5& 671 & $1.6\times10^{-16}$\\
NGC~2264 & 06:41:01.92 09:52:39.0& 352 & $2.9\times10^{-16}$ \\
\enddata
\end{deluxetable*}

The Pa$\alpha$ detection limits for our three sources are given in Table 2 along with the exposure times. Table 3 lists the nearest ionizing source in each cluster along with its estimated spectral type and Lyman continuum luminosity, the projected separation between the head of each tail and this star, the expected flux of Lyman continuum radiation in photons per square centimeter per second, and the predicted flux of Pa$\alpha$ emission under the assumption that each comet tail head is optically thick to Lyman continuum radiation. This estimate ignores the attenuation of the Lyman continuum radiation by dust. Finally, the last column in Table 3 gives the ratio of the observational limit on the Pa$\alpha$ emission derived from Table 2 divided by the expected flux under the assumption that each star with a dust tail contains a hemispherical, high density core that is opaque to the Lyman continuum.

\begin{deluxetable*}{lccccccc}
\tablecolumns{8}
\tabletypesize{\footnotesize}
\tablewidth{0pc}
\tablecaption{Physical properties of the environment of the photoevaporating tails.}
\tablehead{
\colhead{Cluster} &\colhead{O-Star} & \colhead{Spect. Type} & \colhead{Q} &\colhead{d} &\colhead{F(LyC)} &\colhead{$F_{P\alpha}$ predicted} &\colhead{$F_{P\alpha}$ deficit}\\
\colhead{} & \colhead{} & \colhead{ }& \colhead{$\rm \left [ \gamma s^{-1} \right ]$} &\colhead{[cm]} &\colhead{$\rm \left [{\gamma \over {cm^2 s}}\right ]$} &\colhead{$\rm \left [{erg \over {cm^{2}~s~arcsec^{2}}}\right ]$}
}
\startdata
NGC~2244 &HD4615 & O6Ve & $2.2\times 10^{49}$ &$3\times10^{17}$&$1.9\times 10^{13}$&$2.9 \times 10^{-11}$&$1.5\times 10^{-5}$\\
IC~1396 & HD206267 & O6Ve & $2.2\times 10^{49}$ &$1.05\times10^{18}$&$1.6\times 10^{12}$&$2.3 \times 10^{-12}$&$7.0\times 10^{-5}$\\
NGC~2264 & S Mon & O7Ve & $1.3\times 10^{49}$&$9.6\times10^{17}$&$1.1\times 10^{12}$&$1.7 \times 10^{-12}$&$1.7\times 10^{-4}$ \\
\enddata
\end{deluxetable*}

In these estimates, we use the projected distance between the nearest ionizing star and each tail, an assumption that will underestimate the true separation and overestimate the incident Lyman continuum flux. However, we also ignore the contributions of other OB stars in each HII region to the total Lyman continuum radiation field. This assumption will tend to result in an underestimate of the incident radiation field. The estimated incident radiation fields are probably uncertain to about a factor of two. Thus, the predicted emission measures and Pa$\alpha$ fluxes are order-of-magnitude estimates. However, as shown by Table 3, the predicted fluxes are many orders of magnitude higher than the observational limits.

The observations indicate the {\it Spitzer}-detected tails are deficient in Pa$\alpha$ emission by about a factor of one million compared to a simple model in which each dust tail contains a photo-ablating core of dense material. As discussed above, the emission measure, and therefore the fluxes of hydrogen recombination lines are to first order independent of the radius of the ionization front. They depend on the incident Lyman continuum radiation field, and inversely as the distance between the object and source of ionization squared. Thus, a set of optically thick, neutral condensations ought to increase in Pa$\alpha$ flux as the inverse square of the separation. If these dust tails were similar to the proplyds in the Orion Nebula, they would be orders of magnitude brighter than the observed limits.

\subsubsection{The Optically Thin (to Lyman Continuum) Case}

The absence of detectable hydrogen recombination line fluxes might be explained if the disks are optically thin to Lyman continuum radiation and the gas-to-dust ratio is significantly lower than in the ISM. For an assumed radius of the circumstellar environment, $r_I = 10^3$ AU, the critical density above which the gas is opaque to the flux of Lyman continuum radiation is $n_I > (3F/\alpha_B r_I)^{1/2} = 2.8\times 10^4F_{12}^{1/2} r_3^{-1/2}$ where $F_{12}$ is in units of $10^{12}$ Lyman continuum photons ${\rm s^{-1} cm^{-2}}$ and $r_3$ is in units of $10^3$ AU. This implies an emission measure, $EM \approx 6 \times 10^6 F_{12}~{\rm cm^{-6} pc}$. In contrast, our Pa$\alpha$ flux limits (Table~2) imply emission measures of $EM \approx 1673$ ${\rm cm^{-6} pc}$ for the Pa$\alpha$ flux limit from the NGC 2244 tail and $EM \approx 637$ ${\rm cm^{-6} pc}$ for the Pa$\alpha$ fluxes from the tail in IC~1396. Thus, for an incident Lyman continuum  flux $F_{12}$, the  Pa$\alpha$ deficit would be between $2$ and $12 \times 10^{-5}$.  Table~3 gives the actual deficits for each tail given the projected separations from the nearest O-star, the Lyman continuum luminosity of the star (also in Table~3), and the flux limits of Table~2.

If the circumstellar environment is optically thin to the Lyman continuum and has a constant density and radius $r_I$, the upper limit for the Hydrogen mass is $M\approx \mu m_H (11.18 EM[pc])^{1/2}r_I^{5/2}$, which implies masses of $1.7 \times 10^{-6} {\rm M_{\odot}}$ and $2.7 \times 10^{-6} {\rm M_{\odot}}$ for our two limiting cases of EM $\approx$ 637 ${\rm cm^{-6} pc}$ and 1673 ${\rm cm^{-6} pc}$ respectively. These are very conservative upper bounds since observations show that circumstellar environments have flattened disk-like geometries. The emission measure and derived density limits would be an order of magnitude lower for a disk geometry. The mass of a typical protoplanetary disk is about 0.1-0.001 ${\rm M_{\odot}}$ \citep{Andr05}. Our inferred gas mass upper limits for the optically thin case are at least two orders of magnitude lower than the gas masses expected in a typical T Tauri disk, implying gas to dust ratios $10^2$ to $10^4$ times lower than in the ISM . 

\subsection{The Inner Disk}

\citet{Balo06} modeled the grain properties and tail morphologies for the three evaporating disks discussed in this paper.  The models utilized optical constants for astronomical silicates and assumed a single grain size.  It was found that the tail morphologies (faintness at 8 $\mu$m and the apparent lengths) placed strong constraints on the grain sizes. The minimum sizes were of order $\sim$ 0.01 $\mu$m in radius - smaller grains emitted too strongly at 8 $\mu$m to be included in the tails in significant numbers. Grains above $\sim$ 1 $\mu$m in radius tended to be too cold to fit the tail lengths.  The spectra predicted by these models for the objects in IC~1396 and NGC~2264 are shown in Figures \ref{fig:1396IRSSL} and \ref{fig:2264IRSSL} for comparison with our IRS spectra. The fit is satisfactory at wavelengths longer than $\sim$ 12 $\mu$m. At shorter wavelengths, there is significant emission from the "comet head" both from the stellar photosphere and from an excess bridging from 2.4 $\mu$m to 12 $\mu$m and centered on the head. The small difference between the flux of the tail and our model SED at 24 $\mu$m is due to the fact that the slit of LH module of IRS covers the ``head'' and some part of the ``tail'' so the spectra show the flux coming from some combination of the ``head'' and ``tail'' fluxes. We scaled both the spectra and the model SEDs to the flux coming through the slit to make the agreement between the model and the spectra at wavelengths longer than 12 $\mu$m more visible. For the source in NGC~2244, the previous model was modified to improve the fit (see Figure \ref{fig:2244IRSSL}). We used the new 8$\mu$m and 24 $\mu$m fluxes to recalculate the dust mass and the SED for this system. The dust mass in small (0.03 $\mu$m) grains is $3.5 \times 10^{-8} {\rm M_{\sun}}$. We achieved a similarly good fit to the surface brightness distribution with the new mass estimates and the new data that are shown in Fig. 2 of \citet{Balo06}. 

To be conservative, \citet{Balo06} computed dust masses assuming that the mass spectrum extended as $a^{-3.5}$ (where $a$ is the grain size) up to 1mm. The mass goes essentially as the square root of the maximum grain radius in the size spectrum. Anticipating the discussion in Section 4.2, the masses for our models assume a maximum size of $a= 1 ~\mu$m. They are entered in Table 4.

The fits to tail spectra fall far short of accounting for the excess emission above the stellar photospheres in the 3 to 10$\mu$m range, apparent in both our spectra and IRAC photometry. We conclude that these excesses indicate the presence of class II sources in the ``heads'' of our objects. Further tentative evidence of the presence of a class II source in the ``head'' region is a broad 10 $\mu$m silicate emission feature detected in the spectrum of the object in IC~1396.  In the case of NGC~2244 the S/N of the low resolution spectrum is not enough to detect such a feature confidently. On the other hand in the case of NGC~2264 the SED of the unbound dust is rising very sharply between 10 and 15 $\mu$m and this rise might mask the silicate feature at 10 $\mu$m. 

\subsection{Gas in the Inner Disk}

The presence of an inner compact disk agrees well with the models of \citet{John98,Holl04,Adam04} and \citet{Thro05} which all predict a gas rich inner disk surviving the photoevaporation process for longer than $10^5$ years. The SEDs of the ``comet'' heads presented in Figs. \ref{fig:1396IRSSL}, \ref{fig:2264IRSSL}, and \ref{fig:2244IRSSL} are very similar to the lower quartile of T Tauri disks in the Taurus region presented by \citet{Furl06}.

However the Pa$\alpha$ places stringent limits on the gas content of even a compact disk. If a gas disk exists inside 5 AU (which is the gravitational radius at 10 km/s, the sound-speed in ionized H with a cosmic abundance), its surface will be photo-ionized, forming a gravitationally bound (to the central star) HII region.  It is ionized by the external O star, so its density is given by the Stromgren density.
\begin{equation}
\begin{split}
 n_e = \left [{L(LyC) \over 4 \pi \alpha_B} \right ]^{0.5} r_{II}^{-0.5}  D^{-1}~=\\
~5 \times 10^4 L_{49}^{0.5} M_{\sun}^{-0.5} c_{10} D^{-1}~(cm^{-3})
\end{split}
\end{equation}
\noindent where, $r_{II}$ is the gravitational radius, $r_{II} = GM / c_{II}^2$,    where $c_{II}$ = 10 km/s is  the sound speed in photo-ionized gas, M is the mass of the star in the disk in Solar mass units (1 ${\rm M_{\sun}}$),  $L_{49}$ is the Lyman continuum luminosity of the O star in units of $10^{49}$ photons/sec, and $D$ is the distance between the O star and the disk in units of 1 pc.

Thus, a typical disk corona around a Solar mass star will have an electron density ($n_{e2}$) of order 50,000 per cubic cm.   The emission measure will be $EM = n_{e2} L = 1.2 \times 10^4   {\rm cm^{-6} pc}$ for $L$ = 1AU. This is a lower bound because it assumes that the central star makes no contribution to the Lyman continuum luminosity. This large EM will produce a strong Pa$\alpha$ line that should be detected in the narrow band Pa$\alpha$ images. 

The previous calculation applies only for a spatially resolved disk because the emission measure is relevant to the surface brightness assuming that the disk corona is resolved. In our observations the expected size of the inner disk (5-10 AU) subtends a much smaller angle than the resolution element. To apply the calculations to our data (in the case of NGC~2244) we consider an $r_d$ - 5 AU radius disk, located at $d =$ 0.1pc from an O5 star.  We assume that the distance from the Sun is $D =$ 1500 pc, and that each Lyman continuum photon absorbed by the disk results in the emission of  $f =$ 0.1 Pa$\alpha$ photons.

In photo-ionization equilibrium, the Pa$\alpha$ flux from the disk is
\begin{equation}
F(Pa\alpha)  = {f ~{ L(LyC) \over 4 \pi d^2} ~ \pi r_d^2 \over 4 \pi D^2}
\end{equation}

\noindent This is the flux of intercepted LyC photons, multiplied by $f$, divided by $4\pi D^2$, to give the flux at Earth ($L(LyC) = 10^{49}$, $r_d =$ 5 AU, $d =$ 0.1 pc, $D =$ 1.5 kpc.). The resulting flux in this case is F(Pa$\alpha$)=$1.27 \times 10^{-16}~{\rm erg~cm^{-2}~s^{-1}}$. According to the NICMOS Exposure Time Calculator this flux would provide a signal-to noise ratio around 1 so a disk with a radius of 5 AU would not be detected. However the same calculation for a disk with radius of 10 AU yields SNR$>$4 which would be detectable in the NICMOS images, implying an upper limit between 5 and 10 AU for the radius of the inner gas rich part of the disk. This is in agreement with the photoevaporation models.

\section{Discussion}
\subsection{Mass Loss Rates and Amounts}
As we showed in Section 3.1, the Pa$\alpha$ images and the mid-infrared spectra suggest that the gas-to-dust mass ratio in these systems is at least two orders of magnitude lower than the conventional value of 100:1 that was assumed in \citet{Balo06}, so the flow is essentially gas free. 

The low gas-to-dust mass ratio requires changes in the assumptions made when calculating the mass loss rate of these objects. Instead of the value 10 km/s used in \citet{Balo06} assuming that the gas is dragging the dust, we have to use the terminal velocity assuming that the primary forces on the grains are radiation pressure from the O star and a drag force from particles in the surrounding HII region. We use Eq. B.4 of \citet{Aber02} to estimate the terminal velocity. Our calculations show that it is about an order of magnitude higher (around 200-300 km/s) than the value used by \citet{Balo06}. However the new gas-to-dust mass ratio decreases the total amount of outflowing material by about 2 orders of magnitude leaving us with a mass loss rate about 10 times lower than was originally calculated by \citet{Balo06}. Although the lower limit on the mass loss rate of the dust is an order of magnitude higher ($\approx 10^{-11} - 10^{-12} {\rm M_{\odot}/yr}$), the upper limit has not changed significantly because the change in the assumption of the maximum grain size (from 1mm to 1 $\mu$m; see discussion in Section 4.2) and the changes in the assumption about the outflow velocity and in the gas-to-dust ratio cancel each other (leaving only a factor of three difference) resulting in mass loss upper limits of $\approx 10^{-10} - 10^{-11} {\rm M_{\odot}/yr}$. See Table 4 for the values for each individual case.

\begin{deluxetable*}{lcccccccc}
\tablecolumns{8}
\tabletypesize{\footnotesize}
\tablewidth{0pc}
\tablecaption{Physical properties of the photoevaporating sources}
\tablehead{
\colhead{Cluster} &\colhead{$\Delta$l} & \colhead{D} & \colhead{t} &\colhead{vel. disp} &\colhead{$\rm \dot{M}_{max}$} &\colhead{M$\rm _{lost}$}&\colhead{M$\rm
_{dust-max}$} &\colhead{$\rm \dot{M}_{dust}$}\\
\colhead{} & \colhead{[pc]} & \colhead{[pc]}& \colhead{[yr]} &\colhead{[km/s]} &\colhead{$\rm \left [M_{\odot} yr^{-1} \right ]$} &\colhead{$\rm \left [M_{\odot} \right ]$}&\colhead{$\rm \left[M_{\odot} \right ]$}&\colhead{$\rm \left [M_{\odot} yr^{-1} \right ]$}
}
\startdata
NGC~2244 &0.22 & 0.1 &$7\times10^{4}$& 6.8 &$4.73\times 10^{-7}$&0.016&$2 \times 10^{-7}$&$1.5\times 10^{-11} - 9.3 \times 10^{-11}$\\
IC~1396 & 0.21 & 0.35 & $ 10^{5}$ &3.5$^{\star}$& $1.35 \times 10^{-7}$& 0.012&$5 \times 10^{-8}$&$2.4\times10^{-12} - 2.4\times10^{-11}$\\
NGC~2264 & 0.12 & 0.32 & $10^{5}$&3.5&$1.47 \times 10^{-7}$&0.013&$4 \times 10^{-7}$&$3.4\times10^{-11} - 3.4\times10^{-10}$ \\
\enddata
\tablenotetext{$\star$}{no reliable radial velocity dispersion information available for this cluster so we used a typical value for clusters of similar age}
\end{deluxetable*}

\citet{Stor99} analyzed 10 proplyds and found evidence that the photoevaporative flow is dust and metal poor suggesting that most solids are settled in the disk mid-plane. \citet{Thro05} modeled the evolution of a circumstellar disk in the presence of external UV radiation and suggested that as a result of photoevaporation the outer disk becomes completely gas depleted. It is possible that UV radiation has photoablated the disks for our three sources down to the gravitational radius in ionized gas ($\sim$3-5 AU for 1 ${\rm M_{\odot}}$) so these cometary structures are produced by a highly processed disk that has lost most of its gas and started the dust regeneration process in the midplane of the disk. Radiation pressure from the central O-star and dynamical processes in the circumstellar disk would then be expelling this dust from the disk.

Assuming nearly radial orbits and relative velocities between the O-star and the disk-bearing star equal to the cluster velocity dispersion we can estimate the time the disk-bearing stars have spent in the sphere of influence of the O-star \citep[$\le$ 0.5 pc][]{Balo07} in our three systems. Assuming also that the current distances of the disk-bearing stars to the O-stars are the closest point of their orbit to the O-star (giving the lower limit for the total time spent in the 0.5 pc sphere) we find that they have spent about $7-10 \times 10^4$ years within the sphere of influence. We can estimate how much material they lose until they arrive at their current position based on the photoevaporation rate-distance relation of \citet{Rich98} ($\dot{M_{ph}} = 1.46 \times 10^{-6} {\rm M_{\odot}yr^{-1}}  ({ d \over {10^{17} {\rm cm}}})^{-\alpha}$). We adopted $\alpha=1$ from \citet{Balo07}. We integrated the total mass loss during the time the disk-bearing star spends in the sphere of influence of the O-star. Our calculation showed that in our three systems the total lost disk material is about 0.012-0.016 ${\rm M_{\odot}}$. We present the results in Table 4.

The total lost mass is sufficiently large that a typical protoplanetary disk \citep[$M = 10^{-3}$ to $10^{-2} {\rm M_{\sun}}$][]{Andr05} would be totally photoevaporated during the time it spends in the close vicinity of the O-star. We also calculated what the mass loss rate would be at the current positions of our photoevaporating tails, also based on the formula of \citet{Rich98}. We derived values $1-5 \times 10^{-7} {\rm M_{\sun} yr^{-1}}$ which are a couple of orders of magnitude higher than the current mass loss rates (Table 4). This is not a surprise since most of the material is already gone from the disk and also the mechanism that removes the dust from the disk is entirely different from the classical photoevaporation where the flow is gas rich.

It is interesting to compare the mass loss rates for these objects with that for the Vega debris disk, where \citet{Su05} find an  anomalously high loss rate. Taking their estimated total debris mass of $\approx 3 \times 10^{-3} {\rm M_{\earth}}$ in grains between 1 and 46 $\mu$m in radius and their dwell time of $\approx$ 1000 years for these grains near the star, the mass loss is $\approx 10^{-11} {\rm M_{\sun}/yr}$. The evaporating disks are losing grain mass at similar or higher rates (Table 4). 

\subsection{Origin of ``Comet'' Grains}

There are many theoretical treatments of photoevaporation of protoplanetary disks by hot stars \citep[e.g.][]{John98,Rich00,Scall01,Mats03b,Adam04,Thro05}. In general, these treatments agree qualitatively that the process requires $10^5$ \citep[e.g.][]{Thro05} to $10^7$ yrs \citep[e.g.][]{Adam04} to shrink to a core disk of size 5 $-$ 10 AU that is resistant to further erosion \citep{Thro05,Scall01}. Our results suggest that the time scales toward the short end of this range are likely to be applicable.
The evaporation is powered by dust absorption of UV photons, followed by gas heating primarily by the photoelectric effect, with the resulting outflow of hot gas carrying along small, sub-micron sized dust grains. The larger dust grains will be left in a relatively gas-free environment. \citet{Thro05} show that these grains will be in an unstable configuration that will collapse to the disk mid-plain and that planet growth could be triggered as a result. 

There are other consequences. With the gas removed, its damping effects on grain motions (e.g., circularization of orbits) are also removed. Grains will collide and initiate collisional cascades, that lead to a highly elevated rate of production of small, second-generation grains. These grains can be ejected by photon pressure, either due to photons from the central star or, more likely, those from the O-star. This mechanism is a natural way to produce the large numbers of small grains required in our models of the infrared output of the tails of these sources. An additional driver for a large grain collision rate is the effect of photon pressure from the O-star perturbing the orbits of small grains that nevertheless remain gravitationally bound to the low mass star.

The conditions for photon-pressure blow-out of grains in the disk around the low mass star can be derived as follows. Let $\beta$ be the ratio of photon to gravitational force on a grain in the vicinity of the O star by itself. Photon-pressure-driven blow-out occurs for $\beta$ $>$ 0.5. The equivalent ratio of forces for a grain near the low mass star must include its gravity, resulting in a reduction of $\beta$ approximately in proportion to the ratio of gravitational forces from the two stars at the position of the grain. It is easily shown that the modified ratio has a value of 0.5 at a distance from the low mass (mass $M_2$ ) star given by

\begin{equation}
r = {{\gamma d} \over {\left ( {{2 \beta M_1} \over {M_2}} \right ) ^{0.5}}}, 
\end{equation}

\noindent where $d$ is its distance from the high mass star (mass $M_1$). The term $\gamma$ is of order 1 and depends on where the grain is in its orbit. \citet{Lamy97} have determined values of $\beta$ for grains of various compositions and sizes around a number of stars. The values are roughly similar for a given grain size; grain composition enters as a secondary parameter. We will use typical values in the following discussion. Taking the values for the O9.5 V star $\zeta$ Oph, grains of radius between 0.01 and 1 $\mu$m have $\beta \approx$ 1000 - 10000. We take a typical distance between our high and low mass stars to be $d$ = 0.2 pc, and also assume 40 ${\rm M_{\sun}}$ for the high mass star and 0.6${\rm M_{\sun}}$ for the low mass one. We then find that the O-star photon pressure on such small grains becomes dominant at a distance of $\approx$ 110 AU from the low mass star. Small grains well inside this radius will be held in orbit by the gravitational field of the low-mass star. For larger grains, $\beta$ drops rapidly; typically it is $<$ 50 for 10 $\mu$m radius grains. The corresponding distance for them where photon pressure from the O star overcomes the gravity of the low-mass one is $\approx$ 500 AU, that is, outside the boundary of a typical protoplanetary disk \citep[e.g.,][]{Andr05}. Therefore, the larger grains are likely to remain bound to the low mass star throughout the disk, where they will continue to be ground down by collisions.

That is, the grains making up the ``comet tail'' probably originate from collisional cascades in the outer regions of the circumstellar disk. These regions have previously been cleared of gas by photoevaporation and the remaining solid particles are settling toward the disk mid-plain and starting to assemble into larger bodies (see \citet{Thro05} for details), leading to generation of these grains in a vigorous episode of collisional cascades. However, given typical sizes of protoplanetary disks, this process is likely to be effective only up to a maximum grain size of about a micron radius. The tails are composed primarily of grains of $\approx$ 0.01 $\mu$m to $\approx$ 1 $\mu$m in radius (where the lower limit is set by the model fits to their surface brightness profiles; Balog et al. 2006). We have modified the upper mass limits derived by \citet{Balo06} to reflect the approximate grain size limit of 1 $\mu$m. This approximate limit results from the tendency of larger grains to remain gravitationally bound to the low-mass star. 

\subsection{Timescales for Disk Evaporation}

From the mass loss rates and typical disk masses in Section 4.1, we can estimate an approximate timescale of $10^5 - 10^6$ yrs for which this phenomenon is visible. Therefore, the phenomenon might be quite common, however we see only three cases in our GTO survey of about 20 O stars (there are additional cases around three O-stars in the W5 region, Koenig et al. in preparation) suggesting that this is probably a short lived rather rare phenomenon (only about 1/4 of the observed O-stars have cometary structures in their neighborhood).

\section{Conclusions}
We present HST/NICMOS Pa$\alpha$ images and IRS spectra of cometary structures detected in {\it Spitzer}/MIPS 24 $\mu$m images. We estimate an upper limit to the amount of gas in the comets' disk and tail and find that the gas-to-dust mass ratio is much lower than the value observed in the ISM: the tails are essentially gas free. Using this new observation we are able to constrain the flow velocity and thus the mass loss rate. The new mass loss rates allow us to estimate the timescale on which the phenomenon occurs ($10^5 - 10^6$ yr). The short timescale favors photoevaporation models predicting quick removal of gas from the outer parts of the disk.  The ``comet tails'' are produced from the outer regions of the disks, where larger grains collide at an elevated rate generating second generation dust. These small grains are then ejected due to photon pressure from the nearby O-star. The SED of the sources shows excess emission between 3 and 8 $\mu$m, in agreement with the IRS low resolution spectra. This emission indicates that there is an inner disk that survives the photoevaporation process for longer than $10^5$ years, as predicted by the photoevaporation models.

\acknowledgments

The authors thank the anonymous referee for comments and suggestions which improved the paper. We also thank Robert King for providing the VLT data for the source in NGC~2244.
This work is based on observations made with the Spitzer Space Telescope, which is operated 
by the Jet Propulsion Laboratory, California Institute of Technology, under NASA contract 1407. 
Support for this work was provided by NASA through contract 1255094, issued by JPL/Caltech. 
ZB received support from Hungarian OTKA Grants TS049872 and T049082.

\end{document}